\documentclass[reprint,aps,prl,superscriptaddress,showpacs,preprintnumbers]{revtex4-1}

\usepackage{graphicx}
\usepackage{amsmath,amssymb,amsfonts}

\newcommand{\ket}[1]{\left|#1\right\rangle}

\begin{document}

\title{Three-Dimensional Anderson Localization in Variable Scale Disorder}
\author{W. R. McGehee}
\author{S. S. Kondov}
\author{W. Xu}
\author{J. J. Zirbel}
\altaffiliation[Now at]{AOSense, 767 N. Mary Ave., Sunnyvale, CA 94085}
\author{B. DeMarco}
\affiliation{Department of Physics,  University of Illinois at Urbana-Champaign, Urbana, Illinois 61801, USA}
\email{bdemarco@illinois.edu}

\date{\today}

\begin{abstract}
We report on the impact of variable-scale disorder on 3D Anderson localization of a non-interacting ultracold atomic gas. A spin-polarized gas of fermionic atoms is localized by allowing it to expand in an optical speckle potential. Using a sudden quench of the localized density distribution, we verify that the density profile is representative of the underlying single-particle localized states.  The geometric mean of the disordering potential correlation lengths is varied by a factor of four via adjusting the aperture of the speckle focusing lens.  We observe that the root-mean-square size of the localized gas increases approximately linearly with the speckle correlation length, in qualitative agreement with the scaling predicted by weak scattering theory.
\end{abstract}
\maketitle

Anderson localization (AL) is a striking phenomenon in which destructive interference prevents waves from propagating in a disordered medium.  First theoretically investigated in uncorrelated disordered lattices \cite{Anderson1958}, AL has been observed in experiments on light \cite{Wiersma1997,Sperling2013,Segev2013}, sound \cite{Hu2008}, and atomic quantum matter waves \cite{Roati2008,Billy2008,Kondov2011,Jendrzejewski2012}.  Experiments on ultra-cold atom gases afford control over scattering and disorder properties not easily achieved in conventional systems, and measurements on these systems therefore serve as important tests of theory as the initial conditions, disorder potentials, and final states can be well characterized \cite{Sanchez-Palencia2010a}.

AL in 3D is unique compared with lower dimensions because only states with an energy below a critical energy, the mobility edge, are localized \cite{Abrahams1979}.  The validity of approximations used in certain theoretical approaches, such as self-consistent theory \cite{PhysRevB.22.4666}, used to predict how the properties of Anderson-localized waves depend on microscopic features of the disorder in 3D is unknown \cite{SPEPL}.  In this work, we explore how the characteristic disorder length scale affects the thermally averaged localization length for 3D AL of an ultracold atomic gas.  Theoretical work has addressed how  speckle properties impacts AL in 1D \cite{PhysRevLett.98.210401,PhysRevA.80.023605,Piraud.epjst,PhysRevA.79.063617} and 3D \cite{Yedjour2010,1367-2630-15-7-075007,Skipetrov2008,Kuhn2007,SPEPL}.  In contrast to previous measurements of how localization of light in semiconductor powders depends on the discrete grain size \cite{Wiersma1997}, we employ a continuous, spatially correlated disorder potential formed from optical speckle.

We prepare an ultracold gas of $\left(160\pm30\right)\times10^3$ fermionic $^{40}$K atoms confined in an optical dipole trap \cite{supp} using the methods described in Ref.~\cite{Kondov2011}. The gas is spin polarized in the $\ket{F = 9/2, m_F = 9/2}$ hyperfine state and cooled to $\left(174\pm4\right)$~nK. The atoms are non-interacting because collisions are suppressed by the \textit{p}-wave threshold \cite{DeMarco1999}.  A disorder potential created from optical speckle is imposed on the atoms in the trap (Fig. 1).  The speckle is created by passing a 532~nm laser beam through a holographic diffuser and focusing it using a 1.1 f-number lens \cite{Goodman}.  The speckle light forms a repulsive disordered potential proportional to the light intensity, which varies randomly in space.  The strength of the disorder $\Delta$ is characterized by the average speckle potential energy at the focus of the lens (located at the center of the dipole trap) and is adjusted by controlling the 532~nm laser power.  The speckle, slowly turned on over 200~ms, does not significantly change the density or momentum distribution of the trapped gas.

\begin{figure}
\includegraphics[width=1\columnwidth]{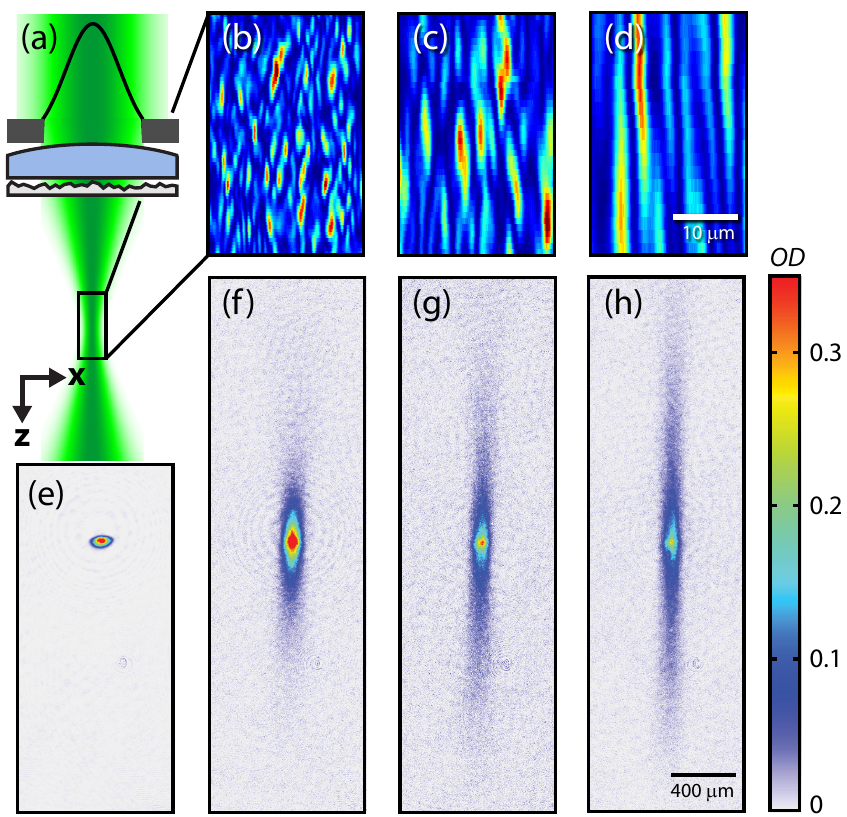}
\label{fig1}
\caption{Optical speckle is created by focusing a 532~nm laser beam (green) through a diffuser (light gray) using a lens (blue), as shown in (a). Truncating the Gaussian beam profile using an aperture (dark gray) controls the speckle correlation lengths, evident in the speckle intensity distributions (b--d) measured via optical microscopy at the focal plane of the lens.  The speckle intensity is shown in false color, with red (blue) regions for high (low) intensity. The density profile of the gas measured by absorption imaging is shown in the dipole trap in (e), and the fully localized profile after expansion into the speckle potential for $\tau=160$~ms in (f--h) for $\Delta=220$~nK$\times k_B$, where $k_B$ is Boltzmann's constant.  The optical depth $OD$ is proportional to the column density.  The speckle correlation length $\overline{\zeta}$=1.2, 2.4, and 3.8~$\mu$m for (b,f), (c,g), and (d,h), respectively.}
\end{figure}

The atoms are localized by suddenly turning off the dipole trap and allowing the gas to expand into the disorder potential.  A magnetic field gradient is applied to cancel the force of gravity during the expansion.  After the density profile of the gas evolves for a time $\tau$, the speckle light is turned off and an absorption image of the column density is obtained.  For $\tau<60$~ms, a two-component distribution is observed, consisting of a mobile component that expands approximately ballistically and a localized part that acquires a static density profile.  Images of the density distribution in the dipole trap and the localized component at long $\tau$ are shown in Fig.~1.

We investigate how varying the correlation length of the speckle potential impacts localization by adjusting the diameter of an aperture in front of the focusing lens.  Speckle is characterized by an intensity autocorrelation with a central feature well described as a cylindrically symmetric Gaussian distribution elongated along the speckle propagation direction $z$ \cite{Gatti2008}.  The size of the Gaussian (i.e., the correlation length of the speckle) is determined by the f-number of the focusing lens.  Reducing the effective diameter of the focusing lens therefore increases the length scale of the speckle, as shown in Fig.~1.  By adjusting the diameter of the aperture from 3.6--13~mm, we vary the $1/e^2$ radii of the autocorrelation from $\zeta_x=0.6$--1.8~$\mu$m in the focal ($x$-$y$) plane and $\zeta_z=3.1$--37~$\mu$m axially. The speckle intensity autocorrelation was directly measured ex-situ \cite{supp}. The corresponding geometric mean $1/e^2$ radius $\overline{\zeta}=\left(\zeta_x^2\zeta_z\right)^{1/3}$, which we use to characterize the correlation length of the disorder, varies from 1.1--4.8~$\mu$m.

As shown in the images in Fig.~1, we observe that the radial ($x$-$y$) size of the localized gas is largely unaffected by changing the speckle correlation length, while the axial ($z$) size grows with increasing $\overline{\zeta}$ (Fig.~1).  To measure the size of the localized gas and systematically investigate the influence of $\overline{\zeta}$, we fit the column density profile to a heuristic function proportional to $e^{-x^2/2\sigma_x^2} e^{-\left | z/\xi_z\right |^\beta}$ with $\sigma_x$, $\xi_z$, and $\beta$ as free parameters.  Because the radial localization length is smaller than the trapped density profile, the in-trap distribution, which has a Gaussian profile, dominates the localized radial distribution.  Along the axial direction, the profile is larger than the axial disorder autocorrelation length and is well described by a stretched exponential.  The stretch exponent $\beta$ enables the fit to evolve from the in-trap distribution ($\beta$=2, Gaussian) to the peaked profiles observed at long $\tau$ ($\beta\leq1$) \cite{supp}.

The measured profiles are a thermal sum over single-particle localized states each of which has an energy dependent, exponentially decaying envelope.  Lacking direct access to the single-particle states, two interpretations of the measured $\xi_z$ at long $\tau$ are possible.  In one case, the single-particle states are small in size with localization lengths comparable to $\zeta_z$.  The density distribution at long $\tau$ is then composed of a collection of small states widely distributed in space, and $\xi_z$ is not representative of the single-particle localization length.  In the second scenario, the localized states are large with sizes comparable to $\xi_z$ and centered on the trapped position of the gas.

To discriminate between these possibilities, we quench the localized density profile at $\Delta=660$~nK$\times k_B$ and $\overline{\zeta}=1.2$~$\mu$m by suddenly removing atoms selectively from the lower half of the gas and measure the resulting dynamics.  Atoms are removed by transferring them to the $\ket{7/2,7/2}$ state via adiabatic rapid passage using a chirped microwave-frequency magnetic field pulse 2~ms in duration that is applied 60~ms after release in the speckle potential.  Spatial selectivity is achieved because the magnetic field gradient applied to compensate gravity shifts the microwave transition frequency.  The $\ket{7/2,7/2}$ atoms exit the imaged field-of-view within 7.5~ms since they experience a downward force greater than gravity.

The inset to Fig.~2a shows the density profile before and after the atoms are removed.  Approximately half of the atoms are removed from the lower ($z>0$) section of the gas.  To quantify the asymmetry of the gas, we alter the fit to have different lengths $\xi_z^<$ and $\xi_z^>$ for $z<0$ and $z>0$, respectively.  As shown in Fig.~2, the quench reduces $\xi_z^>$ by approximately a factor of two while leaving $\xi_z^<$ unperturbed.  The evolution of $\xi_z^>$ and $\xi_z^<$ in time after the quench is shown in Fig. 2.  The atoms reform a symmetric distribution over 300~ms with $\xi_z^>$ and $\xi_z^<$ approximately equal to 90~$\mu$m, which is consistent with the unperturbed profile. This change involves atomic population redistributing in space, as shown in the plot of the ratio $N_</N_>$ in Fig.~2 of the number atoms in the lower and upper halves of the gas (determined from sums of the measured $OD$).  If the single-particle localization lengths were comparable to the 3~$\mu$m speckle autocorrelation length along $z$, then $\xi_z^>$ and $N_</N_>$ would remain static after the quench, since $\xi_z^>$ is much greater than $\zeta_z$.  We therefore interpret the quench dynamics as evidence that the single particle localization lengths are comparable to $\xi_z$, and we define $\xi_z$ as the thermally averaged localization length.

\begin{figure}
\includegraphics[width=1\columnwidth]{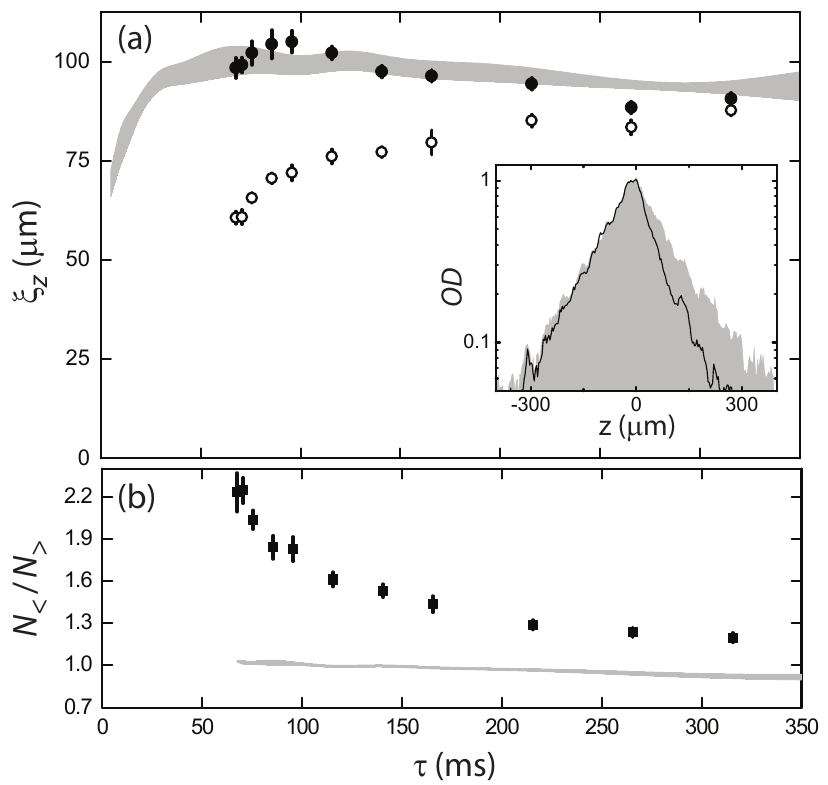}
\label{fig2}
\caption{A slice through the profile of a localized gas is shown in the inset before (gray shaded region) and after (line) removing atoms from the lower half at $\tau=60$~ms.  In (a), the lengths $\xi_z^>$ (open circles) and $\xi_z^<$ (closed circles) are shown relaxing in time after the atoms are removed; the size of the unperturbed gas is shown as a gray band.  The ratio of atoms in the upper and lower half of the gas is shown in (b) for the quenched profile (closed squares) and unperturbed gas (gray band).  The width of the gray bands and the error bars show the standard error in the mean for each point, which is an average of 4--5 measurements.}
\end{figure}

We use the RMS axial size of the gas $\sigma=\xi_z\sqrt{\Gamma\left(1+3/\beta\right)/3\Gamma\left(1+1/\beta\right)}$ to characterize how the localized profile changes as the speckle correlation length is varied ($\Gamma(h)$ is the gamma function with argument $h$). We use $\sigma$ rather than $\xi_z$ because $\beta$ also changes with $\zeta_z$ \cite{supp}.  As shown in Fig.~3 for fixed $\Delta=220$~nK$\times k_B$, $\sigma$ increases with $\tau$ after the trap is turned off until a static distribution is achieved.  The time required to form a stationary profile increases for longer speckle correlation lengths, from approximately 50~ms at $\overline{\zeta}=1.2$~$\mu$m to 150~ms at $\overline{\zeta}=3.8$~$\mu$m.  The rate of expansion during the dynamical period 10~$\mu$m/ms is roughly independent of $\overline{\zeta}$ and corresponds to a 500~nK thermal velocity.  The RMS size of the fully localized component grows from 250~$\mu$m to 1400~$\mu$m for this range of $\overline{\zeta}$.

\begin{figure}
\includegraphics[width=1\columnwidth]{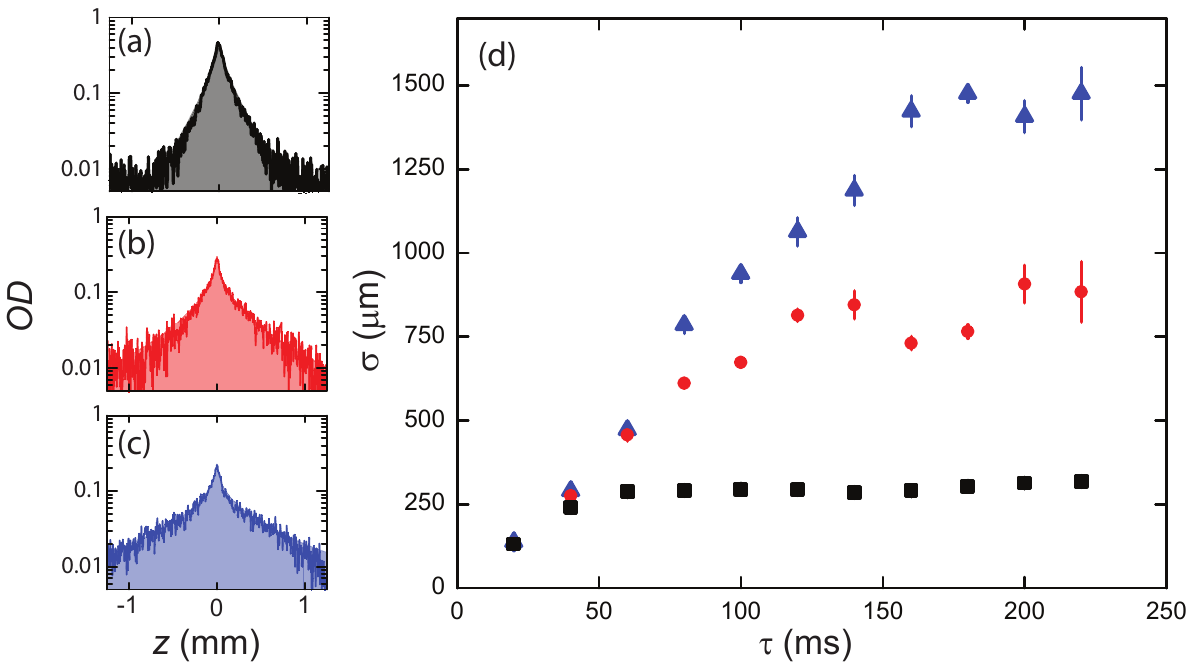}
\label{fig3}
\caption{Representative traces through the center of the localized gas along the $z$-axis for $\overline{\zeta}$=1.2 (a), 2.4 (b), and 3.8~$\mu$m (c) at $\tau$=160~ms. The shaded area shows the heuristic fit to the profile used to determine $\sigma$. In (d), the RMS size of the gas along the axial direction as a function of hold time $\tau$ in the speckle potential is shown for $\overline{\zeta}$ corresponding to that used for profiles (a) (black squares), (b) (red circles), and (c) (blue triangles). Each point is an average of 4--5 measurements and error bars represent the standard error of the mean.}
\end{figure}

The variation in the fully localized $\sigma$ at $\tau\geq$160~ms as $\overline{\zeta}$ is changed is shown in Fig.~4 for fixed $\Delta=220$~nK$\times k_B$.  The RMS size of the gas grows monotonically as $\overline{\zeta}$ is increased.  A fit to a power law $\sigma=A\overline{\zeta}^B$ (solid line in Fig. 4) gives $B=1.22\pm0.06$, indicating approximately linear scaling of $\sigma$ with $\overline{\zeta}$.  This behavior agrees with weak scattering theory for isotropic speckle disorder, which predicts that the single particle localization lengths scale as $\zeta E^2/\Delta^2$ in the limit that the atomic wavevector is larger than the speckle correlation length and for particle energy $E$ is much smaller than the mobility edge \cite{Kuhn2007}.  We infer that distribution of energies in the localized gas is approximately independent of $\overline{\zeta}$ from the fraction of atoms $f$ in the localized component measured at $\tau\geq$160~ms, shown in the inset to Fig.~4.  The localized fraction of atoms is controlled by the distribution of kinetic energies and the mobility edge.  We keep the distribution of kinetic energies in the gas before release into the speckle fixed by holding the temperature of the gas in the trap constant. Since scattering from the disorder is elastic and the localized fraction is invariant with respect to $\overline{\zeta}$, we conclude that the distribution of localized energies $E$ is approximately independent of $\overline{\zeta}$.

\begin{figure}
\includegraphics[width=1\columnwidth]{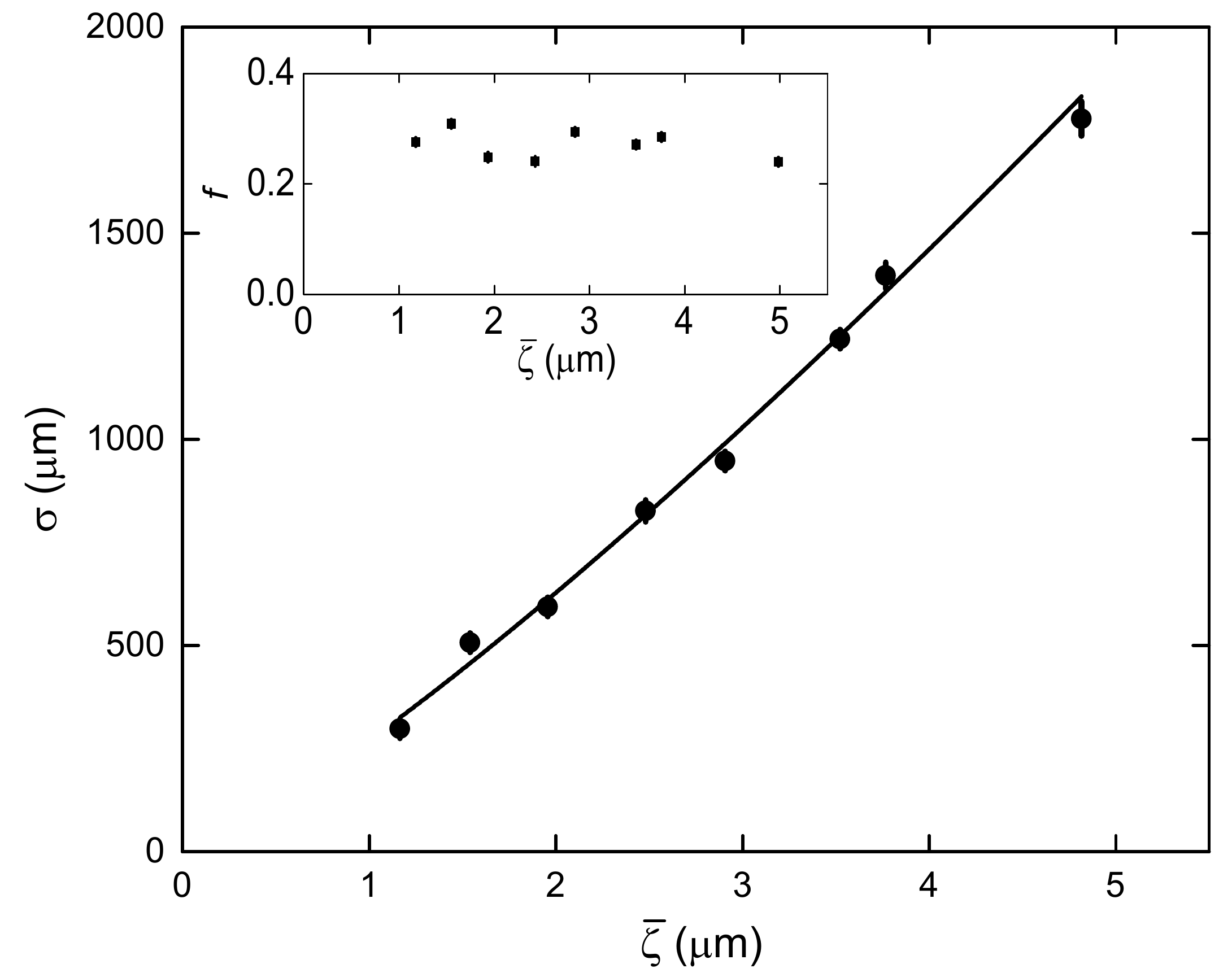}
\label{fig4}
\caption{The RMS axial size of the localized gas $\sigma$ is shown for varied speckle correlation lengths $\overline{\zeta}$ at $\tau$=160~ms. The localized fraction of atoms for each data point is shown in the inset.  Each point is an average of 20 measurements, and the error bars are the standard error in the mean.}
\end{figure}

The magnitude of the observed localization length disagrees with numerical predictions based on self-consistent weak scattering theory for 3D AL in anisotropic speckle potentials \cite{SPEPL}.  This discrepancy may be due to a failure of the weak scattering approximation at the onset of localization.  The work in Ref.~\cite{SPEPL} also neglects the variation in the speckle intensity arising from the speckle envelope \cite{supp}.  In the future, digital optical holography \cite{Pasienski:08,Gaunt:12} may be used to explore how different types of optical disorder affect the localization length and mobility edge in 3D.

\begin{acknowledgements}
The authors acknowledge financial support the Army Research Office and National Science Foundation and thank Laurent Sanchez-Palencia and Marie Piraud for helpful discussions.
\end{acknowledgements}

\bibliographystyle{prsty}
\bibliography{wrm_references}

\begin{thebibliography}{10}

\bibitem{Anderson1958}
P.~W. Anderson, Phys. Rev. {\bf 109},  1492  (1958).

\bibitem{Wiersma1997}
D.~S. Wiersma, P. Bartolini, A. Lagendijk, and R. Righini, Nature {\bf 390},
  671  (1997).

\bibitem{Sperling2013}
T. Sperling, W. B\"{u}hrer, C. Aegerter, and G. Maret, Nat. Photon. {\bf 7},
  48  (2013).

\bibitem{Segev2013}
D.~C. M.~Segev, Y.~Silberberg, Nat. Phot. {\bf 7},  197  (2013).

\bibitem{Hu2008}
H. Hu {\it et~al.}, Nat Phys {\bf 4},  945  (2008).

\bibitem{Roati2008}
G. Roati {\it et~al.}, Nature {\bf 453},  895  (2008).

\bibitem{Billy2008}
J. Billy {\it et~al.}, Nature {\bf 453},  891  (2008).

\bibitem{Kondov2011}
S.~S. Kondov, W.~R. McGehee, J.~J. Zirbel, and B. DeMarco, Science {\bf 334},
  66  (2011).

\bibitem{Jendrzejewski2012}
F. Jendrzejewski {\it et~al.}, Nat Phys {\bf 8},  398  (2012).

\bibitem{Sanchez-Palencia2010a}
L. Sanchez-Palencia and M. Lewenstein, Nat Phys {\bf 6},  87  (2010).

\bibitem{Abrahams1979}
E. Abrahams, P.~W. Anderson, D.~C. Licciardello, and T.~V. Ramakrishnan, Phys.
  Rev. Lett. {\bf 42},  673  (1979).

\bibitem{PhysRevB.22.4666}
D. Vollhardt and P. W\"olfle, Phys. Rev. B {\bf 22},  4666  (1980).

\bibitem{SPEPL}
M. Piraud, L. Pezzé, and L. Sanchez-Palencia, EPL {\bf 99},  50003  (2012).

\bibitem{PhysRevLett.98.210401}
L. Sanchez-Palencia {\it et~al.}, Phys. Rev. Lett. {\bf 98},  210401  (2007).

\bibitem{PhysRevA.80.023605}
P. Lugan {\it et~al.}, Phys. Rev. A {\bf 80},  023605  (2009).

\bibitem{Piraud.epjst}
M. Piraud and L. Sanchez-Palencia, Eur. Phys. J. - Spec. Top. {\bf 217},  91
  (2013).

\bibitem{PhysRevA.79.063617}
E. Gurevich and O. Kenneth, Phys. Rev. A {\bf 79},  063617  (2009).

\bibitem{Yedjour2010}
A. Yedjour and B.~A. Van~Tiggelen, Eur. Phys. J. D {\bf 59},  249  (2010).

\bibitem{1367-2630-15-7-075007}
M. Piraud, L. Pezzé, and L. Sanchez-Palencia, New J. Phys. {\bf 15},  075007
  (2013).

\bibitem{Skipetrov2008}
S.~E. Skipetrov, A. Minguzzi, B.~A. van Tiggelen, and B. Shapiro, Phys. Rev.
  Lett. {\bf 100},  165301  (2008).

\bibitem{Kuhn2007}
R.~C. Kuhn {\it et~al.}, New J. Phys. {\bf 9},  161  (2007).

\bibitem{supp}
See Supplemental Material at [] for information on experimental parameters, the
  speckle autocorrelation, and the stretched exponential fit.

\bibitem{DeMarco1999}
B. DeMarco {\it et~al.}, Phys. Rev. Lett. {\bf 82},  4208  (1999).

\bibitem{Goodman}
J. Goodman, {\em Speckle Phenomena in Optics} (Roberts and Company, Englewood,
  CO, 2007).

\bibitem{Gatti2008}
A. Gatti, D. Magatti, and F. Ferri, Phys. Rev. A {\bf 78},  063806  (2008).

\bibitem{Pasienski:08}
M. Pasienski and B. DeMarco, Opt. Express {\bf 16},  2176  (2008).

\bibitem{Gaunt:12}
A. Gaunt and Z. Hadzibabic, Sci. Rep. {\bf 2},  721  (2012).

\end{thebibliography}
\end{document}